\documentclass{article}

\usepackage{arxiv}

\usepackage[numbers, compress]{natbib}

\usepackage[utf8]{inputenc} % allow utf-8 input
\usepackage[T1]{fontenc}    % use 8-bit T1 fonts
\usepackage{hyperref}       % hyperlinks
\usepackage{url}            % simple URL typesetting
\usepackage{booktabs}       % professional-quality tables
\usepackage{amsfonts}       % blackboard math symbols
\usepackage{nicefrac}       % compact symbols for 1/2, etc.
\usepackage{microtype}      % microtypography
\usepackage{xcolor}         % colors

\usepackage{multirow}
\usepackage{adjustbox}
\usepackage{amsmath}

\newcommand{\para}[1]{\noindent{\bf #1}}

\title{OpenHEXAI: An Open-Source Framework for Human-Centered Evaluation of Explainable Machine Learning}

\date{} 					% Or removing it

\author{
Jiaqi Ma\thanks{Equal contribution.}\\
UIUC
\And
Vivian Lai\footnote[1]{} \\
Visa Research
\And
Yiming Zhang\\
Carnegie Mellon University
\And
Chacha Chen\\
University of Chicago
\And
Paul Hamilton\\ 
Harvard University
\And
Davor Ljubenkov\\
Harvard University
\And 
Himabindu Lakkaraju\thanks{Equal supervision.}\\
Harvard University
\And
Chenhao Tan\footnote[2]{}\\
University of Chicago
}

\renewcommand{\headeright}{ }
\renewcommand{\undertitle}{ }

\begin{document}

\maketitle

\begin{abstract}
Recently, there has been a surge of explainable AI (XAI) methods driven by the need for understanding machine learning model behaviors in high-stakes scenarios. However, properly evaluating the effectiveness of the XAI methods inevitably requires the involvement of human subjects, and conducting human-centered benchmarks is challenging in a number of ways: designing and implementing user studies is complex; numerous design choices in the design space of user study lead to problems of reproducibility; and running user studies can be challenging and even daunting for machine learning researchers. To address these challenges, this paper presents OpenHEXAI, an open-source framework for human-centered evaluation of XAI methods. OpenHEXAI features (1) a collection of diverse benchmark datasets, pre-trained models, and post hoc explanation methods; (2) an easy-to-use web application for user study; (3) comprehensive evaluation metrics for the effectiveness of post hoc explanation methods in the context of human-AI decision making tasks; (4) best practice recommendations of experiment documentation; and (5) convenient tools for power analysis and cost estimation.
OpenHEAXI is the first large-scale infrastructural effort to facilitate human-centered benchmarks of XAI methods. It simplifies the design and implementation of user studies for XAI methods, thus allowing researchers and practitioners to focus on the scientific questions. Additionally, it enhances reproducibility through standardized designs. Based on OpenHEXAI, we further conduct a systematic benchmark of four state-of-the-art post hoc explanation methods and compare their impacts on human-AI decision making tasks in terms of accuracy, fairness, as well as users' trust and understanding of the machine learning model.
\end{abstract}

\section{Introduction}
\label{sec:intro}

Explainable Machine Learning/Artificial Intelligence (XAI) methods aim to provide \emph{human-understandable} insights on machine learning models. These insights are crucial in high-stakes applications such as hiring, loan approvals, or medical diagnosis, where the algorithmic decisions made by machine learning models can significantly impact people's lives. Although there has been a surge in the development of XAI methods in recent years, it is highly non-trivial to evaluate and compare these methods properly. Unlike measuring the accuracy of AI predictions, quantifying explainability objectively is difficult since it is fundamentally dependent on human interpretation, and there is still little consensus on a specific set of objective metrics suitable for measuring explainability. As a result, human-centered evaluation (user study) is often adopted, which typically involves soliciting feedback from users who interact with an XAI method and then analyzing the feedback data to investigate how well the method is providing human-understandable information. However, designing and conducting user studies can be complicated and expensive, and lead to incomparable outcomes across different studies. To date, there has not been a systematic benchmark for human-centered evaluation of XAI methods. In this work, we propose OpenHEXAI, an Open-source framework for human-centered Evaluation of XAI methods, aiming to address the aforementioned challenges and establish a systematic and replicable benchmark. 

Designing and conducting user studies for evaluating XAI methods can be challenging for a few reasons. Firstly, one can measure the explainability of an XAI method from a diverse lens. For example, does the XAI method improve the performance of human-AI collaboration? Does the XAI method improve human trust in AI predictions? Does the human truly understand why the AI predicted these predictions? 
Different questions often require different carefully designed user studies to answer them properly. Secondly, conducting a user study requires non-trivial efforts in implementing the infrastructure, such as web applications and user interface (UI) designs. Thirdly, there are many design choices in the user study setup. From the UI design to the selection of the user population, each design choice may have an influence on the study results. The numerous design choices significantly increase the difficulty of reproducing user studies. Last but not least, user studies are challenging to manage and time-consuming since they require human labor.

The proposed OpenHEXAI framework is carefully designed to address the above challenges. Firstly, we conduct a literature survey on human-centered evaluation of post hoc explanation methods in order to better navigate through the existing design space of user studies in this area. In particular, we have designed the framework in the context of human-AI joint decision making~\citep{Lai2021-fl} and evaluate post hoc explanation methods in terms of the joint decision performance. Secondly, we develop an open-source software for implementing user studies, including a web application with different UI configurations, 7 benchmark datasets with trained machine learning models on each dataset, 6 state-of-the-art post hoc explanation methods, and a comprehensive set of evaluation metrics covering accuracy, trust, and fairness of the human-AI decision. The software is highly extensible and can be used as a boilerplate to accommodate new datasets, models, post hoc explanation methods, and UI designs. We have also included instructions to deploy user studies on Prolific\footnote{https://www.prolific.co/}. 
Thirdly, motivated by data cards and model cards~\citep{Gebru2021-dn,McMillan-Major2021-cl}, we summarize an \emph{evaluation card} for human-centered evaluation of XAI, which outlines a checklist of best practices of user study designs for the evaluation of XAI. Finally, the OpenHEXAI framework also provides tools for power analysis and cost estimates to assist researchers in better budgeting for user studies. 

Utilizing the proposed OpenHEXAI framework, we further conduct a systematic benchmark for four state-of-the-art post hoc explanation methods on two datasets. Through both objective evaluation metrics and subjective survey questions, we evaluate the post hoc explanation methods in terms of their impacts on the efficacy and fairness of the human-AI joint decisions, as well as user's trust and understanding of the AI system. 

Overall, our OpenHEXAI framework provides the first large-scale infrastructural effort to streamline human-centered benchmarks for XAI methods. The framework is easily extensible and reusable by researchers and practitioners to carry out user studies of new XAI methods on new datasets and pre-trained models. In addition, we will keep updating our software library to enhance the usability and capability of OpenHEXAI.

\section{Literature Survey}
\label{sec:related}

In order to understand the infrastructure needs of researchers for evaluating XAI methods, we first carry out a literature survey. We focus on literature that conducts user studies for post hoc explanation methods. This survey was carried out near the end of 2022, leading to a summary of 17 papers with publication dates spanning between 2013 and 2022. In this survey, we examine the papers from various perspectives, including task scenarios of the user study, post hoc explanation methods used, and major research questions posed. While this survey aims to be thorough, it does not comprehensively cover the existing literature, given the wide range of research questions related to human interactions with post hoc explanation methods. The primary goal of the survey is to understand researchers' interests and needs in human-centric benchmarking of post hoc explanation methods.

\paragraph{Task Scenarios.} The papers in this survey cover a variety of task scenarios of user studies for understanding the impact of post hoc explanations. One group of studies~\citep{Nguyen2018-fr,Zhang2020-sj,Liu2021-rg,Kiani2020-dg,fan2021humanai,Hadash2022,Weerts2019-cw,Bansal2021-sb} focus on the human-AI decision making scenario where humans make decisions with the assistance of AI. The specific tasks include both making predictions on benchmark classification datasets that only require common sense, such as news classification or income prediction~\citep{Nguyen2018-fr,Zhang2020-sj}, and tasks that require domain expertise, such as medical diagnosis~\citep{Kiani2020-dg}. Another group of work~\citep{Chromik2021-un,Chandrasekaran2018-gi,Hase2020-cg,Colin2022-mn} tries to gauge whether and how post hoc explanation methods improve humans' understanding of the AI system by asking the subjects to predict the AI's prediction, which is also commonly known as forward simulation. There are also a few less commonly used task scenarios, such as model debugging~\citep{Balayn2022-ko,chen2022usecasegrounded} or counterfactual reasoning~\citep{chen2022usecasegrounded}.

\paragraph{Post hoc Explanation Methods.} Existing user studies cover a rather narrow set of post hoc explanation methods in comparison to the rich literature in this area. Among the papers in this survey, SHAP~\citep{Lundberg2017-yq} is the most commonly used method~\citep{Zhang2020-sj,Weerts2019-cw,Chromik2021-un,zhou2022exsum,Hadash2022,chen2022usecasegrounded}, with LIME~\citep{ribeiro2016should} closely following~\citep{Nguyen2018-fr,Hase2020-cg,zhou2022exsum,Hadash2022,chen2022usecasegrounded}. Other methods covered include Vanilla Gradients~\citep{simonyan2013deep}, Gradient $\times$ Input (GI)~\citep{ancona2017towards}, SmoothGrad (SG)~\citep{smilkov2017smoothgrad}, Integrated Gradients (IG)~\citep{sundararajan2017axiomatic}, Occlusion~\citep{zeiler2014visualizing}, CAMs~\citep{zhou2016learning}, Grad-CAM~\citep{selvaraju2017grad}, Anchors~\citep{ribeiro2018anchors}, etc.

\paragraph{Research Questions.} The research questions posed in each paper are fairly diverse so it is difficult to enumerate them succinctly. We summarize a few common research questions based on the task scenarios. In the context of human-AI decision making, common research questions include whether explanations improve the human decision accuracy~\citep{Weerts2019-cw,Nguyen2018-fr,Liu2021-rg,Kiani2020-dg,Zhang2020-sj}, the level of understanding of how the AI system works~\citep{Hadash2022}, human trust in the AI system~\citep{Zhang2020-sj}, etc. In some studies, other elements that complement explanations are also considered, such as confidence scores~\citep{Zhang2020-sj}, positive framing of explanations~\citep{Hadash2022}, the type of users (with different levels of expertise)~\citep{Kiani2020-dg}, or psychological mechanisms of users~\citep{yang2022psychological}. In the context of forward simulation, how explanations improve human understanding of the AI system is a major research question shared by multiple studies~\citep{Chromik2021-un,Chandrasekaran2018-gi,Hase2020-cg,Colin2022-mn}. 

\section{The OpenHEXAI Framework}
\label{sec:overview}

In this section, we present an overview of the proposed OpenHEXAI framework. We first introduce the application scenario in which the benchmark is situated. Then we introduce the three core modules of the open-source software: a machine learning module equipped with various ready-to-use datasets, pre-trained models, and post hoc explanation methods; a web application module that includes a web user interface and the corresponding backend server for user studies; and an evaluation module capable of automatically calculating various objective evaluation metrics based on the user responses in the user study. The modular design of our open-source software allows researchers to reuse part of the software flexibly. Furthermore, we outline an evaluation card, providing recommendations for documenting the specifics of user study design and execution in evaluating XAI methods. Finally, we provide convenient tools for power analysis and cost estimates of user studies.

\subsection{Benchmark Scenario}

There are a variety of application scenarios of XAI, and the effectiveness of XAI methods can be highly dependent on the target scenario. Therefore, in order to have meaningful comparisons across different XAI methods, we ground the OpenHEXAI framework in the context of a specific type of application scenario, human-AI joint decision making~\citep{Lai2021-fl}.

In this scenario, the AI models assist human subjects by providing predictions or recommendations for certain decision making tasks, such as online content moderation or medical diagnosis. XAI methods can be applied to improve the transparency of the assistive models, potentially leading to better human-AI joint decisions.

The human-AI joint decision making scenario presents an apt context for benchmarking and comparing different XAI methods. Firstly, this scenario is a canonical application scenario for XAI methods with many practically relevant applications. Additionally, the human-AI joint decision making scenario is a well-defined context with specific needs and constraints. The efficacy of the XAI methods can be objectively measured in terms of the quality of the joint decisions. This clarity helps reduce ambiguity in the design and execution of benchmarks, which is crucial for reproducibility. Narrowing the focus to this specific application scenario allows us to develop a common ground benchmark where researchers can compare and replicate each other's results.

\subsection{Machine Learning Module}

The machine learning module is responsible for preparing the data instances, model predictions, and feature attribution scores calculated by post hoc explanation methods for the downstream web application (see Section~\ref{sec:webapp}) that directly interacts with users in the user study. We build on top of an existing XAI benchmark library, OpenXAI~\citep{agarwal2022openxai}\footnote{OpenXAI only provides evaluation metrics that can be automatically calculated from the data and explanations, as opposed to human-based evaluation centered in the proposed OpenHEXAI framework.}, while adding an additional layer to make the datasets, models, and post hoc explanation methods in OpenXAI ready to use for user study. 

Firstly, we implement wrappers for a collection of datasets, pre-trained models, and post hoc explanation methods. To enable a simple and uniform API, we create abstractions such as \texttt{Instance}, \texttt{Dataset}, and \texttt{Explainer} that are compatible with any tabular dataset as well as post hoc explanation methods. In addition to the existing datasets and models in OpenXAI, OpenHEXAI further processes two other datasets and pre-train models on them to enrich the application domains. We have also written codebooks for each dataset to explain the semantics of features for users properly. Overall, OpenHEXAI compiles a collection of real-world datasets in diverse high-stakes domains, including finance, criminal justice, and education. Each dataset comes with two pre-trained machine learning models, one logistic regression model, and one neural network model. Table~\ref{tab:datasets} summarizes the datasets available in OpenHEXAI. For each dataset and their pre-trained models, OpenHEXAI further supports 6 state-of-the-art post hoc explanation methods inherited from OpenXAI: LIME~\citep{ribeiro2016should}, SHAP~\citep{Lundberg2017-yq}, Vanilla Gradients (Grad)~\citep{simonyan2013deep}, Gradient $\times$ Input (GI)~\citep{ancona2017towards}, SmoothGrad (SG)~\citep{smilkov2017smoothgrad}, and Integrated Gradients (IG)~\citep{sundararajan2017axiomatic}.

At the deployment time, researchers can create a configuration file to specify the choice of a dataset, machine learning model, and post hoc explanation attribution method to be used in a user study. After initializing the dataset and model as configured, the machine learning module randomly samples test instances and computes model predictions as well as feature attribution scores. The instances, predictions, and feature attribution scores are stored in a NoSQL database for the web application to query during the user study.

\begin{table}[h]
    \centering
    \caption{Datasets available in OpenHEXAI. RCDV and Student Admission are the newly processed datasets that did not exist in OpenXAI. The protected attribute refers to a characteristic that is legally shielded from discrimination.}
    \label{tab:datasets}
    \begin{adjustbox}{width=\textwidth}
    \begin{tabular}{l l r r l}
    \toprule
       Domain & Dataset Name & Data Size & Feature Dimension & Protected Attribute \\
       \midrule
       \multirow{5}{*}{Finance} & German Credit~\citep{Dua:2019} & 1,000 & 60 & Gender \\
        & HELOC~\citep{FICO} & 9,871 & 23 & - \\
        & Adult Income~\citep{yeh2009comparisons} & 45,222 & 13 & Gender \\
        & Give Me Some Credit~\citep{GiveMe} & 102,209 & 10 & - \\
        \midrule
       \multirow{2}{*}{Criminal Justice} & RCDV~\citep{schmidt1988predicting} & 9,549 & 16 & Gender, Ethnicity \\
        & COMPAS~\citep{jordan15:effect} & 6,162 & 7 & Gender, Ethnicity \\
        \midrule
       Education & Student Admission~\citep{cheng2019explaining} & 100 & 29 & - \\
    \bottomrule
    \end{tabular}
    \end{adjustbox}
\end{table}

\subsection{Web Application Module}
\label{sec:webapp}

Now we introduce the web application module that interacts with users.

\begin{figure*}[t]
    \vskip -10pt
    \centering
    \includegraphics[width=0.85\textwidth]{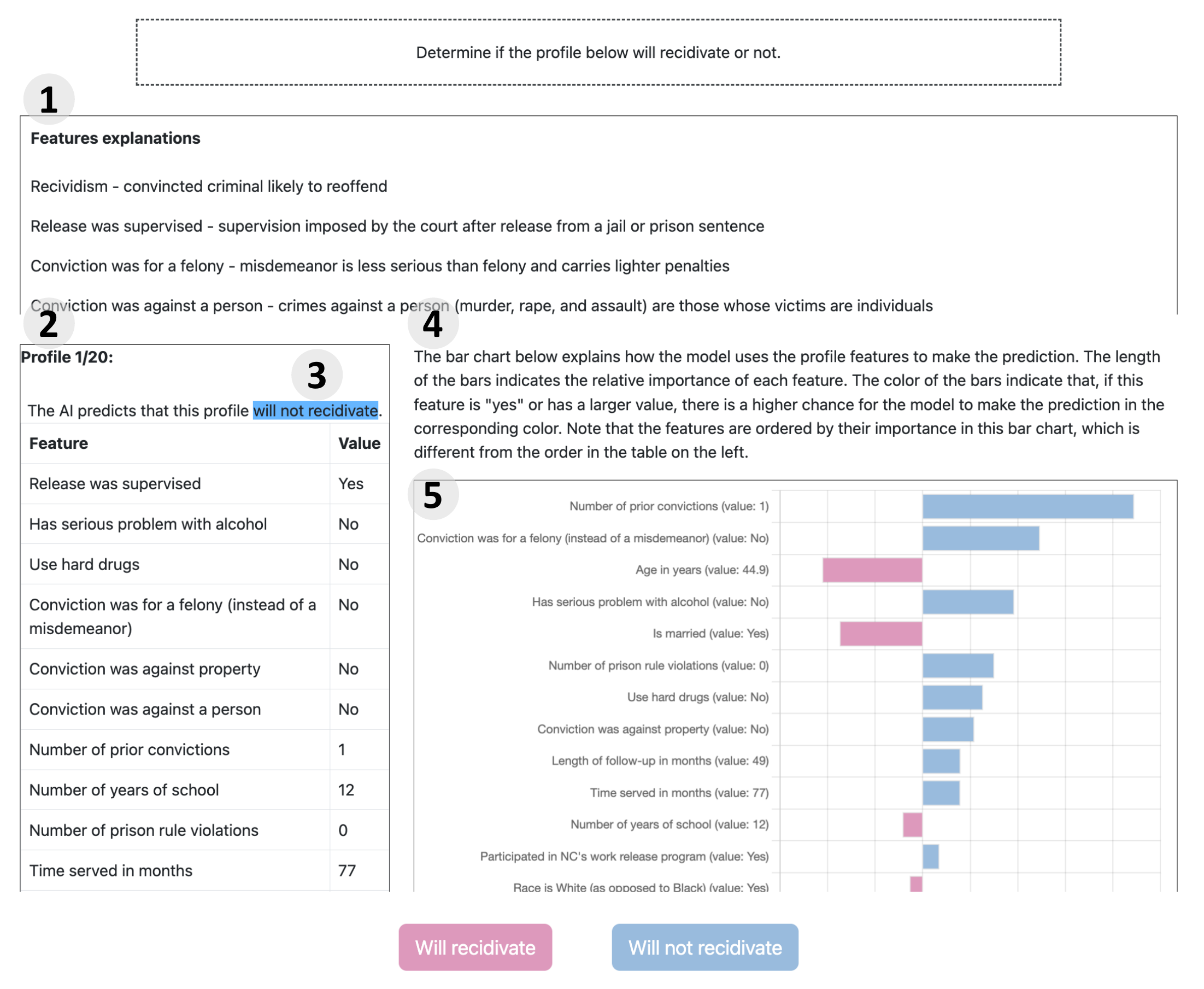}
    \caption{This figure illustrates the task page for the RCDV dataset and conditions with the predicted label and explanations. (1) shows a box including explanations for features that require more explanations. (2) shows the profile of a defendant. (3) shows the predicted label. (4) is a description of how the bar chart could be interpreted. Finally, (5) shows the bar chart that orders features by their absolute feature importance scores.}
    \label{fig:task_page}
    \vskip -10pt
\end{figure*}

\paragraph{User Study Task Flow.} All participants go through four phases on the web application during the user study: (1) consent page; (2) instructions and attention check; (3) task phase; (4) exit survey. The purpose of the attention check questions is to ensure that the participants read the instructions before attempting the user study. For example, we added True or False questions about the purpose of the study and the AI assistance that they will be provided, if any, as attention-check questions. Participants who fail the attention check are disqualified. During the task phase, each participant completes 20 predictions. A screenshot of the task phase interface is shown in Figure~\ref{fig:task_page}. In the exit survey, besides the set of Likert scale questions (see Section~\ref{sec:evaluation-module}), we also collect basic demographic information. 

\paragraph{Adaptable and Reusable Design.} The web application module is designed to be easily adaptable and reusable. Different datasets and post hoc explanation methods can be easily configured through a main configuration file without the researcher editing the web application code. The main configuration file takes in a few important arguments, such as \texttt{dataset} and \texttt{explainer config}. Additionally, the HTML templates could be customizable and reused by other datasets by redefining them in the configuration file. The attention checks and exit survey questions could also be easily changed by editing the JSON files. Currently, our module can be extended to tabular datasets, and we plan to include text and image datasets in the future. Please refer to Appendix~\ref{app:webapp} for more details.

\subsection{Evaluation Module}
\label{sec:evaluation-module}

The evaluation module is a script that reads the user study results and automatically calculates various objective evaluation metrics. The evaluation module currently supports the following metrics. \texttt{Accuracy} and \texttt{F1}: the classification accuracy and F1 score of the human decision or the AI prediction against the ground truth label in the dataset. \texttt{AVG Time}: the average time spent by the user to process an instance. \texttt{Over-Reliance}: the proportion of instances where the human adopts the AI prediction while the AI prediction is incorrect. \texttt{Under-Reliance}: the proportion of instances where the human does not adopt the AI prediction while the AI prediction is correct. \texttt{AAOD} (Average Absolute Odds Difference): $\frac{1}{2} [|\text{FPR}_{A=\text{minority}} - \text{FPR}_{A=\text{majority}}| + |\text{TPR}_{A=\text{minority}} - \text{TPR}_{A=\text{majority}}|]$, where FPR and TPR are respectively the false positive rate and true positive rate, and $A$ is the protected attribute. \texttt{EOD} (Equal Opportunity Difference): $|\text{TPR}_{A=\text{minority}} - \text{TPR}_{A=\text{majority}}|$. \texttt{AAOD} and \texttt{EOD} respectively measures the degree of unfairness of binary classification decisions under the notion of Equalized Odds and Equal Opportunity~\citep{barocas2017fairness}.

\subsection{Evaluation Card}
Detailed documentation of the development of machine learning techniques has gained significant attention recently~\citep{Gebru2021-dn,McMillan-Major2021-cl}. In particular, there have been several documentation guidelines on documenting the data collection process and the machine learning model training process, such as data cards and model cards~\citep{McMillan-Major2021-cl}. As both the datasets and machine learning models continue to grow in complexity, there is an increasing number of ad hoc decisions being made in the data collection and model training processes. These decisions may implicitly bias the model development and potentially result in harmful consequences when the datasets or models are used in unsuitable contexts. Data cards and model cards complement conventional academic papers or technical reports, advocating for detailed documentation to enhance the transparency of the machine learning model development.

The human-centered evaluation of XAI methods shares similarities with the data collection and model training processes. Numerous ad hoc decisions must be made during the user study design. Some of the decision choices may be largely irrelevant to the main hypothesis under investigation and are too tedious to be included in the paper. Furthermore, researchers developing the XAI methods may not be familiar with the process of documenting user studies. Motivated by data cards and model cards, we propose an evaluation card that outlines a checklist for documenting user studies under the OpenHEXAI framework. 

\paragraph{The design of the evaluation card.} We organize the evaluation card by three phases of the user study: design phase, execution phase, and analysis phase. For each phase, we list a set of questions asking for detailed information about the user study. This evaluation card serves as an initial attempt to document the complex human-centric evaluation of XAI. It will be further improved as we conduct more user studies using this framework.

Design phase:
\begin{enumerate}
    \item Is the study pre-registered? If so, what is the link to the pre-registration?
    \item What is the budget for this study?
    \item How are the decision making tasks/datasets chosen?
    \item What is the target population?
    \begin{enumerate}
        \item What are the exact inclusion criteria for the study?
        \item Does the study focus on experts in a given domain or lay people?
    \end{enumerate}
    \item Is there an attention check for the study?
    \item What efforts have been made to help the user understand the semantic meaning of the task and data?
    \item What are major design considerations for the user interface? What are alternative design choices that are not used?
\end{enumerate}

Execution phase:
\begin{enumerate}
    \item Is there a pilot study? What adjustments are made after pilot study?
    \item What is the compensation rate for the participants?
\end{enumerate}

Analysis phase:
\begin{enumerate}
    \item Are there any participants excluded from the analysis? What are the exclusion criteria?
\end{enumerate}

An example evaluation card of our benchmark study (Section~\ref{sec:exp}) can be found in Appendix~\ref{app:evaluation-card}.

\subsection{Power Analysis and Cost Estimate}
Conducting research with human subjects can be costly. Therefore, it is crucial to estimate the necessary number of participants for a user study and the total cost of the study. OpenHEXAI provides reference numbers for these estimates. 

Since we have run a benchmark study on selected datasets available in OpenHEXAI (see Section~\ref{sec:exp}), we are able to conduct power analyses based on the user responses collected in our benchmark studies. The number of participants required to achieve statistically significant conclusions for a particular experiment on a dataset depends on both the noise or complexity level of the dataset and the specific experimental setup (e.g., the choice of explanation methods for comparison and the evaluation metrics). While the power analysis derived from our benchmark study does not directly apply to future experiments designed by researchers using OpenHEXAI, it could still serve as a useful reference given the shared datasets and web application interface. Specifically, we carry out a power analysis for ANOVA on the mean accuracy of several groups of human-AI joint decisions. In different groups, the users are given model explanations by different post hoc explanation methods.

Additionally, we provide a script to estimate the cost for a given configuration of benchmark study. This is calculated on the basis of both the power analysis result and the average time spent by users in our benchmark study. Due to space limit, more details are provided in Appendix~\ref{app:power-analysis}.
\section{A Benchmark Study Based on OpenHEXAI}
\label{sec:exp}
In this section, we conduct a benchmark study using our OpenHEXAI framework. 

\paragraph{Research Questions.}

As discussed in Section~\ref{sec:related}, researchers have been using user studies to investigate multiple aspects of XAI methods in human-AI joint decision making applications. First, we can use user studies to measure \emph{whether} XAI methods improve the performance of the human-AI teams, which could be measured by both the accuracy of the join decisions and the efficiency of the joint decision making process. In addition, user studies can provide us with deeper insights about \emph{how} XAI methods influence the joint decision making process. For example, do the humans trust or understand their AI teammate's decisions? Can humans spot potential biases in their AI teammate? In this study, we compare different post hoc explanation methods by answering the following research questions with respect to each post hoc explanation method.

\begin{enumerate}
    \item RQ1 (Efficacy): Do post hoc explanations of AI predictions improve the accuracy and/or efficiency of joint human-AI predictions?
    \item RQ2 (Trust): How do post hoc explanations of AI predictions affect human's trust in AI? Do they lead to overtrust or undertrust of AI?
    \item RQ3 (Understanding): Do post hoc explanations of AI predictions improve human's understanding on how the models work?
    \item RQ4 (Fairness): Do post hoc explanations of AI predictions improve the fairness of joint human-AI predictions?
\end{enumerate}

\paragraph{Experiment Configuration.} We focus on a selection of datasets and post hoc explanation methods available in OpenHEXAI for this benchmark study. Specifically, we conduct user studies on two datasets, German Credit and RCDV, which are two commonly used datasets in the XAI literature. We evaluate four post hoc explanation methods, LIME, SHAP, SG, and IG, by applying them to the pre-trained neural network model on each dataset. Each post hoc explanation method corresponds to one user study condition, where we provide the user with the data features, model prediction, and the post hoc feature attributions, and ask the user to make a final decision. Additionally, we include two control conditions, one with only data features, and another with data features and model prediction. For writing convenience, we denote the data feature only condition as \textbf{F}, the data feature and model prediction condition as \textbf{FP}, and the condition with post hoc explanations as \textbf{FPE}. Therefore, each dataset has 6 user study conditions (\textbf{F}, \textbf{FP}, \textbf{FPE-LIME}, \textbf{FPE-SHAP}, \textbf{FPE-SG}, and \textbf{FPE-IG}), and there are 12 user study conditions in total.  There are approximately 30 users recruited for each condition, totaling 371 participants. The exact number of participants in each condition is reported in Appendix~\ref{app:num-participants}.
Each user will make decisions on 20 data points, followed by an exit survey. For each dataset, we first randomly draw 200 test data points from the whole test set, and then for each user, we randomly draw 20 data points from the 200 data points. The same 200 test data points are used across all the user study conditions on each dataset.

\begin{table}
    \centering
    \vskip -10pt
    \caption{Objective metrics of human(-AI) decisions. Error bars stand for standard error of the mean.}
    \label{tab:objective_metrics_results}
    \begin{adjustbox}{width=\textwidth}
        \begin{tabular}{l l r r r r r r r}
        \toprule
           Dataset & Condition & Accuracy & F1 & Avg Time & Over-Reliance & Under-Reliance & AAOD & EOD \\
           \midrule
           \multirow{6}{*}{\begin{tabular}[c]{@{}l@{}}German\\ credit\end{tabular}} 
            & F & 0.497$\pm$0.02 & 0.596$\pm$0.02 & 6.47$\pm$0.59 & 0.192$\pm$0.02 & 0.312$\pm$0.02 & 0.103 & 0.155 \\ % german-ann-lime_1
            
            & FP & 0.624$\pm$0.02 & 0.752$\pm$0.02 & 5.93$\pm$0.71 & 0.234$\pm$0.02 & 0.143$\pm$0.01 & 0.118 & 0.171 \\ % german-ann-lime_2
            
            & FPE-LIME & 0.602$\pm$0.02 & 0.731$\pm$0.02 & 8.03$\pm$2.99 & 0.269$\pm$0.02 & 0.129$\pm$0.01 & 0.174 & 0.175 \\ % german-ann-lime_3
            
            & FPE-SHAP & 0.758$\pm$0.02 & 0.835$\pm$0.02 & 3.73$\pm$0.5 & 0.198$\pm$0.02 & 0.044$\pm$0.01 & 0.167 & 0.288 \\ % german-ann-shap_3
            
            & FPE-SG & 0.552$\pm$0.02 & 0.69$\pm$0.02 & 5.6$\pm$0.82 & 0.29$\pm$0.02 & 0.158$\pm$0.01 & 0.168 & 0.117 \\ % german-ann-sg_3
            
            & FPE-IG & 0.737$\pm$0.02 & 0.812$\pm$0.02 & 4.98$\pm$0.73 & 0.168$\pm$0.02 & 0.095$\pm$0.01 & 0.076 & 0.141 \\ % german-ann-ig_3
            \midrule
           \multirow{6}{*}{RCDV} 
            & F & 0.533$\pm$0.02 & 0.433$\pm$0.02 & 4.58$\pm$0.57 & 0.219$\pm$0.02 & 0.248$\pm$0.02 & 0.032 & 0.01 \\ % rcdv-ann-lime_1
            
            & FP & 0.545$\pm$0.02 & 0.405$\pm$0.02 & 4.17$\pm$0.53 & 0.318$\pm$0.02 & 0.137$\pm$0.01 & 0.13 & 0.148 \\ % rcdv-ann-lime_2
            
            & FPE-LIME & 0.58$\pm$0.02 & 0.406$\pm$0.02 & 3.57$\pm$0.45 & 0.288$\pm$0.02 & 0.132$\pm$0.01 & 0.084 & 0.093 \\ % rcdv-ann-lime_3
            
            & FPE-SHAP & 0.57$\pm$0.02 & 0.467$\pm$0.02 & 4.92$\pm$0.42 & 0.29$\pm$0.02 & 0.14$\pm$0.01 & 0.071 & 0.11 \\ % rcdv-ann-shap_3
            
            & FPE-SG & 0.528$\pm$0.02 & 0.381$\pm$0.02 & 4.88$\pm$0.54 & 0.336$\pm$0.02 & 0.136$\pm$0.01 & 0.08 & 0.053 \\ % rcdv-ann-sg_3
            
            & FPE-IG & 0.568$\pm$0.02 & 0.443$\pm$0.02 & 4.83$\pm$0.99 & 0.283$\pm$0.02 & 0.148$\pm$0.01 & 0.053 & 0.083 \\ % rcdv-ann-ig_3
        \bottomrule
        \end{tabular}
    \end{adjustbox}
\end{table}

\begin{table}
    \centering
    \caption{Objective metrics of the AI predictions.}
    \label{tab:test_set_objective}
        \begin{tabular}{l r r r r}
        \toprule
           Dataset & Accuracy & F1 & AAOD & EOD \\
           \midrule
           \multirow{1}{*}{German credit} 
            & 0.675 & 0.806 & 0.148 & 0.296 \\
           \multirow{1}{*}{RCDV} 
            & 0.595 & 0.391 & 0.088 & 0.1 \\
        \bottomrule
        \end{tabular}
        \vskip -10pt
\end{table} 

\paragraph{Evaluation Metrics.}

To quantitatively answer the aforementioned research questions, we use a mix of objective metrics and subjective survey questions.

The objective metrics have been introduced in Section~\ref{sec:evaluation-module}. For efficacy (RQ1), we use \texttt{Accuracy} and \texttt{F1} to measure the accuracy of the human-AI joint decision while using \texttt{Avg Time}, which calculates the average time taken for one prediction to measure the efficiency. For trust (RQ2), we measure it by \texttt{Over-Reliance} and \texttt{Under-Reliance}. For fairness (RQ4), we report \texttt{AAOD} and \texttt{EOD}. For understanding (RQ3), we mainly rely on subjective survey questions.

In the exit survey, we include a set of Likert scale questions, adapted from existing literature. Each question requires users to choose from five given options: \textit{Strongly Disagree}, \textit{Disagree}, \textit{Neutral}, \textit{Agree}, and \textit{Strongly Agree}. In our subsequent analysis, these choices are mapped onto a scale of 1 to 5, with 5 representing \textit{Strongly Agree}. The mean scores of responses for each question are then reported. The survey questions are listed below, each annotated with the corresponding research question that informs its design, as well as the literature where we adapt the questions from.
\begin{itemize}
    \item Q1 (RQ2): I am confident and comfortable using the system's predictions~\cite{anik2021data}.
    \item Q2 (RQ2): I trust the system's predictions~\cite{anik2021data}.
    \item Q3 (RQ4): The system has integrity~\cite{anik2021data}.
    \item Q4 (RQ2): The system's predictions are sound based on the explanations provided~\cite{anik2021data}.
    \item Q5 (RQ3): I am confident about my decisions~\cite{green_disparate_2019}.
    \item Q6 (RQ2): The system's predictions influenced my decisions~\cite{green_disparate_2019,van_berkel_effect_2021}.
    \item Q7 (RQ3): I understand the process by which the prediction was made~\cite{binns2018s}.
    \item Q8 (RQ3): I understand how the model works to predict whether [outcome will be 0 or 1]~\cite{wang2021explanations}.
    \item Q9 (RQ3): I can predict how the model will behave~\cite{wang2021explanations}.
    \item Q10 (RQ3): The explanations are clear~\cite{green_disparate_2019}.
    \item Q11 (RQ3): The explanations are easy to understand~\cite{guo_visualizing_2019}.
    \item Q12 (RQ3): The explanations give me ideas about the data used in the system~\cite{anik2021data}.
    \item Q13 (RQ3): The explanations are informative~\cite{anik2021data}.
    \item Q14 (RQ4): Fair models do not systematically discriminate against individuals with a common ethnicity, religion, gender, etc., irrespective of whether the relevant group enjoys legal protections. Based on the above definition of fairness, this model is fair~\cite{binns2018s}.
    \item Q15 (RQ4): I could identify and correct unfair predictions made by the model~\cite{van_berkel_effect_2021}.
    \item Q16 (RQ4): I had all the required information to assess the fairness of the model~\cite{van_berkel_effect_2021}.
\end{itemize}
All questions appear in conditions with post hoc explanations, while none of them appear in condition \textbf{F}. Q4 and Q10 - Q13 do not appear in the condition \textbf{FP}.

\paragraph{Results of Objective Metrics.} The objective metrics of the human-AI joint decisions (or human decisions in condition \textbf{F}) are reported in Table~\ref{tab:objective_metrics_results}. As a reference, the objective metrics of AI prediction alone are reported in Table~\ref{tab:test_set_objective}. While AI assistance, i.e., predictions and post hoc explanations, generally do not have a significant effect on human-AI joint performances in RCDV in terms of \texttt{Accuracy} (53\% to 58\%), human-AI joint performances (55\% to 76\%) are significantly better than human decisions (50\%) in German credit. LIME, SHAP, and IG generally improve human-AI joint performance more than SG. Remarkably, human-AI joint performance, when equipped with SHAP or IG, achieves better performance than both human alone and AI alone, which is not achieved by the \textbf{FP} condition without post hoc explanations.

Different post hoc methods have a different effect on \texttt{Over-Reliance} on German credit as compared to RCDV. For example, while SHAP explanations result in the lowest \texttt{Over-Reliance} on German credit, it results in the highest \texttt{Over-Reliance} on RCDV. On the other hand, LIME explanations have a similar effect on \texttt{Over-Reliance} in both datasets.

LIME explanations reduce \texttt{AAOD} and \texttt{EOD} in German credit and RCDV, compared to only showing predictions. However, \texttt{AAOD} and \texttt{EOD} become worse when additional information is provided. We hypothesize that explanations generally highlight and unintentionally magnify unnecessary and unimportant features, thus hindering humans from making good judgments on fairness.

\begin{table*}
    \centering
    \vskip -10pt
    \caption{Results of subjective survey questions Q1-Q4. M: mean. SD: standard deviation.}
    \label{tab:subjective_metrics_results}
    \begin{adjustbox}{width=0.9\textwidth}
        \begin{tabular}{l l r r r r }
        \toprule
           Dataset & Condition & Q1 & Q2 & Q3 & Q4 \\
           \midrule
           \multirow{5}{*}{\begin{tabular}[c]{@{}l@{}}German\\ credit\end{tabular}}
            & FP & M=3.33, SD=1.22 & M=3.27, SD=1.11 & M=3.58, SD=0.78 & - \\
            & FPE-LIME & M=3.57, SD=0.92 & M=3.57, SD=0.92 & M=3.53, SD=0.81 & M=3.63, SD=0.91 \\
            & FPE-SHAP & M=3.35, SD=1.06 & M=3.42, SD=1.13 & M=3.42, SD=0.87 & M=3.52, SD=1.01 \\
            & FPE-SG & M=3.60, SD=1.14 & M=3.53, SD=1.12 & M=3.60, SD=1.02 & M=3.60, SD=1.08 \\
            & FPE-IG & M=3.00, SD=1.26 & M=2.90, SD=1.17 & M=3.13, SD=0.96 & M=3.10, SD=1.22 \\
            \midrule
           \multirow{5}{*}{RCDV} 
            & FP & M=3.32, SD=0.96 & M=3.29, SD=0.96 & M=3.29, SD=0.92 & - \\
            & FPE-LIME & M=3.27, SD=1.03 & M=3.07, SD=1.03 & M=3.30, SD=0.78 & M=3.33, SD=0.94 \\
            & FPE-SHAP & M=3.23, SD=0.96 & M=3.00, SD=1.00 & M=2.90, SD=0.87 & M=3.37, SD=0.98 \\
            & FPE-SG & M=3.35, SD=1.06 & M=3.42, SD=1.13 & M=3.42, SD=0.87 & M=3.52, SD=1.01 \\
            & FPE-IG & M=3.27, SD=1.18 & M=3.10, SD=1.14 & M=3.17, SD=0.86 & M=3.17, SD=1.13 \\
        \bottomrule
        \end{tabular}
    \end{adjustbox}
    \vskip -10pt
\end{table*}

\begin{table*}
    \centering
    \caption{Results of subjective survey questions Q5-Q8. M: mean. SD: standard deviation.}
    \label{tab:subjective_metrics_results2}
    \begin{adjustbox}{width=0.9\textwidth}
        \begin{tabular}{l l r r r r }
        \toprule
           Dataset & Condition & Q5 & Q6 & Q7 & Q8 \\
           \midrule
           \multirow{5}{*}{\begin{tabular}[c]{@{}l@{}}German\\ credit\end{tabular}} 
            & FP & M=3.61, SD=1.18 & M=3.67, SD=1.09 & M=3.18, SD=1.40 & M=2.97, SD=1.31 \\
            & FPE-LIME & M=3.53, SD=1.02 & M=4.13, SD=0.96 & M=3.97, SD=0.88 & M=3.63, SD=0.95 \\
            & FPE-SHAP & M=3.65, SD=1.00 & M=3.84, SD=1.14 & M=3.39, SD=1.04 & M=3.39, SD=1.10 \\
            & FPE-SG & M=3.63, SD=1.05 & M=4.00, SD=1.10 & M=3.53, SD=1.20 & M=3.33, SD=1.16 \\
            & FPE-IG & M=3.40, SD=1.14 & M=3.43, SD=1.12 & M=3.20, SD=1.19 & M=3.00, SD=1.16 \\
            \midrule
           \multirow{5}{*}{RCDV} 
            & FP & M=3.48, SD=1.07 & M=3.42, SD=1.16 & M=3.00, SD=1.19 & M=3.10, SD=1.20 \\
            & FPE-LIME & M=3.57, SD=1.02 & M=3.60, SD=0.88 & M=3.73, SD=1.00 & M=3.43, SD=1.0 \\
            & FPE-SHAP & M=3.77, SD=1.02 & M=3.57, SD=0.96 & M=3.80, SD=0.75 & M=3.70, SD=0.74 \\
            & FPE-SG & M=3.75, SD=0.87 & M=3.53, SD=1.17 & M=3.57, SD=1.06 & M=3.50, SD=1.15 \\
            & FPE-IG & M=3.90, SD=0.87 & M=3.27, SD=1.21 & M=3.43, SD=0.99 & M=3.47, SD=1.02 \\
        \bottomrule
        \end{tabular}
    \end{adjustbox}
    \vskip -10pt
\end{table*}

\begin{table*}
    \centering
    \caption{Results of subjective survey questions Q9-Q12. M: mean. SD: standard deviation.}
    \label{tab:subjective_metrics_results3}
    \begin{adjustbox}{width=0.9\textwidth}
        \begin{tabular}{l l r r r r }
        \toprule
           Dataset & Condition & Q9 & Q10 & Q11 & Q12 \\
           \midrule
           \multirow{5}{*}{\begin{tabular}[c]{@{}l@{}}German\\ credit\end{tabular}} 
            & FP & M=2.65, SD=1.06 & - & - & - \\
            & FPE-LIME & M=3.07, SD=0.85 & M=3.90, SD=0.70 & M=3.97, SD=0.92 & M=4.23, SD=0.72 \\
            & FPE-SHAP & M=3.07, SD=1.05 & M=3.58, SD=0.87 & M=3.52, SD=1.10 & M=3.71, SD=0.96 \\
            & FPE-SG & M=3.30, SD=1.10 & M=3.43, SD=1.31 & M=3.40, SD=1.28 & M=3.90, SD=0.94 \\
            & FPE-IG & M=2.87, SD=1.20 & M=3.20, SD=1.11 & M=3.27, SD=1.18 & M=3.37, SD=1.14 \\
            \midrule
           \multirow{5}{*}{RCDV} 
            & FP & M=2.65, SD=1.06 & - & - & - \\
            & FPE-LIME & M=3.03, SD=0.88 & M=3.73, SD=0.85 & M=3.83, SD=0.86 & M=3.90, SD=0.87 \\
            & FPE-SHAP & M=2.77, SD=0.99 & M=3.77, SD=0.80 & M=3.93, SD=0.57 & M=3.90, SD=0.54 \\
            & FPE-SG & M=3.00, SD=1.15 & M=3.72, SD=1.13 & M=3.84, SD=0.94 & M=4.00, SD=0.94 \\
            & FPE-IG & M=2.87, SD=1.09 & M=3.67, SD=0.98 & M=3.70, SD=0.97 & M=4.00, SD=0.89 \\
        \bottomrule
        \end{tabular}
    \end{adjustbox}
    \vskip -10pt
\end{table*}

\begin{table*}[h!]
    \centering
    % \vskip -15pt
    \caption{Results of subjective survey questions Q13-Q16. M: mean. SD: standard deviation.}
    \label{tab:subjective_metrics_results4}
    \begin{adjustbox}{width=0.9\textwidth}
        \begin{tabular}{l l r r r r }
        \toprule
           Dataset & Condition & Q13 & Q14 & Q15 & Q16 \\
           \midrule
           \multirow{5}{*}{\begin{tabular}[c]{@{}l@{}}German\\ credit\end{tabular}} 
            & FP & - & - & M=3.30, SD=1.09 & M=3.24, SD=1.26 \\
            & FPE-LIME & M=4.03, SD=0.71 & M=3.50, SD=1.23 & M=3.30, SD=0.86 & M=3.67, SD=0.91 \\
            & FPE-SHAP & M=3.71, SD=1.11 & M=3.42, SD=1.16 & M=3.58, SD=0.91 & M=3.52, SD=0.76 \\
            & FPE-SG & M=3.63, SD=1.02 & M=3.23, SD=1.38 & M=3.63, SD=0.88 & M=3.77, SD=0.72 \\
            & FPE-IG & M=3.53, SD=1.09 & M=3.53, SD=1.20 & M=3.13, SD=1.20 & M=3.23, SD=0.96 \\
            \midrule
           \multirow{5}{*}{RCDV} 
            & FP & - & - & M=3.03, SD=1.12 & M=2.81, SD=1.12 \\
            & FPE-LIME & M=3.83, SD=0.86 & M=3.27, SD=0.96 & M=3.50, SD=0.85 & M=3.37, SD=0.95 \\
            & FPE-SHAP & M=4.10, SD=0.65 & M=2.93, SD=1.12 & M=3.27, SD=0.81 & M=3.23, SD=1.06 \\
            & FPE-SG & M=3.91, SD=0.95 & M=3.63, SD=1.08 & M=3.34, SD=0.99 & M=3.59, SD=1.11 \\
            & FPE-IG & M=3.50, SD=0.99 & M=3.53, SD=0.81 & M=3.60, SD=0.99 & M=3.93, SD=0.93 \\
        \bottomrule
        \end{tabular}
    \end{adjustbox}
    \vskip -12pt
\end{table*}

\paragraph{Results of Subjective Survey Questions.} The user responses to the exit survey questions are summarized in Table~\ref{tab:subjective_metrics_results},~\ref{tab:subjective_metrics_results2},~\ref{tab:subjective_metrics_results3}, and~\ref{tab:subjective_metrics_results4}. The results for subjective metrics generally support our observations in objective metrics. Q1, Q2, Q4, and Q6 are questions about trust in the model and its predictions. The results suggest that participants generally have the most trust in LIME explanations as opposed to the rest. When participants are shown LIME and SHAP explanations, results from Q7, Q8, Q11, and Q13 suggest that participants tend to understand better the model and how the predictions are made. Lastly, while LIME explanations reduce fairness, as shown in the results of objective metrics, results from Q14, Q15, and Q16 seem otherwise. In general, the type of post hoc explanations does not help gauge the fairness of the model accurately.

\section{Limitations}
We discuss a few limitations of the proposed benchmark framework to better articulate the practical applicability of this framework.

Firstly, while we have deliberately incorporated datasets across different domains, there is still room to further improve the quantity and diversity of the supported datasets. We will include more datasets as we continue to develop this framework.

Secondly, this framework is designed for a specific application scenario of XAI, i.e.,  human-AI decision making, in order to create a reproducible benchmark suite. Furthermore, we have limited our focus to post hoc explanation methods, for two main reasons. First, post hoc explanation methods are arguably among the most popular types of XAI methods. Second, this focus simplifies the API design for the benchmark. The investigation of more general XAI methods or other application scenarios is beyond the scope of this framework.

Thirdly, while we have collected a wide range of objective and subjective evaluation metrics measuring four critical aspects of XAI methods, we have yet to fully understand the implications and compatibility of these metrics. In particular, several pressing questions warrant further investigation for future studies: (1) How stable/noisy are these metrics? (2) Are different metrics measuring the same aspect compatible with each other? (3) If they are not compatible, what criteria should guide the selection of appropriate evaluation metrics? As we accumulate more benchmark study results, the variance and correlation among these metrics may provide insights into these questions. However, being able to evaluate these metrics in a reproducible fashion is an essential preliminary step for addressing more complex scientific inquiries concerning them, which is the main focus of this study.

Lastly, while the proposed evaluation card approach holds great promise for enhancing the reproducibility of human-centric benchmarks of XAI methods, we acknowledge that our current implementation could be further improved to encompass a broader range of user study design elements. For a more detailed discussion on potential improvements, please refer to Appendix~\ref{app:eval-card-future}.

\section{Conclusion}
\label{sec:conclusion}

This paper has introduced OpenHEXAI, an innovative open-source framework designed to address the challenges of human-centered evaluations of XAI methods. The framework offers an integrated solution that simplifies the design and execution of user studies, thus allowing researchers and practitioners to focus on scientific aspects of the study and enhancing reproducibility through standardized designs. Additionally, we have utilized OpenHEXAI to perform a systematic benchmark study of four state-of-the-art post hoc explanation methods, LIME, SHAP, SmoothGrad, and Integrated Gradients, comprehensively investigating how these methods impact the efficacy and fairness of the human-AI joint decision, as well as user's trust and understanding of AI. With an array of features including diverse benchmark datasets, pretrained models, post hoc explanation methods, a user-friendly web application, comprehensive evaluation metrics, and best practice guidelines, OpenHEXAI can promote a wider adoption of human-centered evaluation of XAI methods and accelerate research in this field.

\bibliographystyle{plainnat}
\bibliography{reference}

\newpage
\appendix
\section{Web Application}
\label{app:webapp}

The web application has been thoughtfully designed to allow researchers and industry practitioners to download the repository and conduct user studies with minimal modifications required. A key feature facilitating this blend of customizability and user-friendliness is the configuration file. It allows users to specify various parameters such as the dataset, post hoc explanation method, and model.

Given that different datasets may involve diverse prediction tasks, the text and instructions on the HTML files can be adjusted accordingly. Additionally, modifications to the attention check questions and survey inquiries can be swiftly made via the JSON file located in the respective dataset folders.

While our web application currently supports only tabular datasets, we have plans in the pipeline to expand its capabilities to accommodate text and image datasets, further enhancing its utility and reach.

The source code of the web application is hosted at \url{https://github.com/ChicagoHAI/OpenHEXAI}.

\section{Evaluation Card}

\subsection{The Evaluation Card for Our Benchmark Study}
\label{app:evaluation-card}

\para{Design phase:}

\para{1.} Yes, the study was pre-registered at \url{https://aspredicted.org/g5et5.pdf}.

\para{2.} The budget for this study was 2,000 US dollars.

\para{3.} We chose the datasets based on a few factors. Firstly, we intended to choose popular datasets as this is a benchmark study. Secondly, we tried to diversify the domains. Finally, we decided the number of datasets based on our budget.

\para{4 (a).} We used the following three criteria to filter participants on Prolific: (1) residing in the United States; (2) English is their first language; (3) minimal approval rate of 95\%.

\para{4 (b).} This study focuses on lay people as it is intended to serve as a benchmark study that can be easily reproducible.

\para{5.} Yes, there is a simple attention check with two yes and no questions asking about the purpose of the study.

\para{6.} We included an instruction on the decision making task at both the beginning of the user study and each task page. We further gave explanations to data features that have complicated semantic meanings.

\para{7.} One major design consideration was about how to visualize the feature attribution scores returned by the post hoc explanation methods. As shown in Figure~\ref{fig:task_page}, we used bar chart to visualize the relative importance of the feature attributions, and added an instruction on top of the bar chart about how to interpret it. Another major design consideration is the order of the features. For the feature table, we manually ordered the features based on the semantic meaning of the features. For the feature attribution bar chart, we ordered the features by the absolute attribution scores. An alternative design was to use the same feature order for the table and the bar chart. But we ended up using different feature orders as they are respectively more intuitive for the table and the bar chart.

\para{Execution phase:}

\para{1.} Yes, there was a pilot study on 10 participants to test the workflow. We did not make any adjustments after the pilot study as everything looked good.

\para{2.} The compensation rate was 9.92 US dollars per hour.

\para{Analysis phase:}

\para{1.} No participants were excluded from the analysis.

\subsection{Future Directions to Improve the Evaluation Card} 
\label{app:eval-card-future}

Admittedly, the evaluation card outlined in this work has not comprehensively covered all aspects of user study design. In this section, we discuss a few directions where we plan to continue expanding the evaluation card. 

\paragraph{Dataset and model preparation.} When new datasets are introduced, the first aspect we plan to address revolves around the preparation of datasets and models. Questions to include in this category might be:

\para{1.} What are the basic statistics of the dataset, including size, diversity, and feature distribution?

\para{2.} How was the training setup for the models determined, including architecture, hyperparameters, and optimization techniques?

\para{3.} How was the test data selected for the user study? What sampling techniques were used?

\para{4.} Were the datasets balanced or imbalanced? How did this affect the study?

\para{5.} Did the datasets undergo any preprocessing or augmentation?

\paragraph{User interface design.} The second aspect focuses on enriching the questions related to user interface design. Example questions that can be added to this category include:

\para{1.} How many features are being visualized on a single page?

\para{2.} What mechanisms are used to generate feature explanations?

\para{3.} How was the color scheme determined? Does it follow any specific guidelines or theories (e.g., color theory)?

\para{4.} Is the user interface designed to be accessible to individuals with disabilities?

\para{5.} How are interactive elements like buttons or sliders integrated and what is their purpose?

\paragraph{Survey question design.} The third aspect aims to further delve into the design of survey questions. We propose adding the following example questions:

\para{1.} On average, how much time is required to answer each survey question?

\para{2.} What types of questions are included in the survey (e.g., Likert scale, open-ended, closed-ended)?

\para{3.} Are there redundant questions embedded in the survey to check for answer quality or consistency?

\para{4.} Is there a logical flow or grouping of questions in the survey?

\para{5.} How are potentially sensitive or personal questions handled? Are there opt-out options?

\section{Power Analysis}
\label{app:power-analysis}

We conduct a power analysis for the two datasets, German Credit and RCDV, using our benchmark study results. Specifically, we perform power analysis for an ANOVA test using the \texttt{Accuracy} results of the 6 experiment conditions. When setting the significance level as 0.05 and the power as 0.8, we obtain that the required sample size for German Credit is 154, while the required sample size for RCDV is 22,395. 

These numbers cannot be directly used for future user studies, as the hypotheses to be tested and the expeirment settings might vary. However, it becomes clear that conducting user study on RCDV gives much noisier results in comparison to German Credit. So these numbers provide a good reference for future researchers who wish to use our framework. 

We will continue to provide such references as we conduct user studies on more datasets.

\section{Benchmark Study Details}
\label{app:benchmark}

\subsection{Exact Number of Participants}
\label{app:num-participants}
The exact numbers of participants are listed in Table~\ref{tab:participants_per_condition}.
\begin{table}[h]
    \centering
    \caption{Number of participants for each condition.}
    \label{tab:participants_per_condition}
        \begin{tabular}{l l r}
        \toprule
           Dataset & Condition & Number of Participants \\
           \midrule
           \multirow{6}{*}{\begin{tabular}[c]{@{}l@{}}German\\ credit\end{tabular}} 
            & F & 30 \\
            & FP & 34 \\
            & FPE-LIME & 31 \\
            & FPE-SHAP & 31 \\
            & FPE-SG & 30 \\
            & FPE-IG & 30 \\
            \midrule
           \multirow{6}{*}{RCDV} 
            & F & 32 \\
            & FP & 31 \\
            & FPE-LIME & 30 \\
            & FPE-SHAP & 30 \\
            & FPE-SG & 32 \\
            & FPE-IG & 30 \\
        \bottomrule
        \end{tabular}
\end{table}

\subsection{Task Page Interface}
\label{app:task-page-interface}

\begin{figure*}[t]
    \centering
    \includegraphics[width=0.8\textwidth,angle=-90]{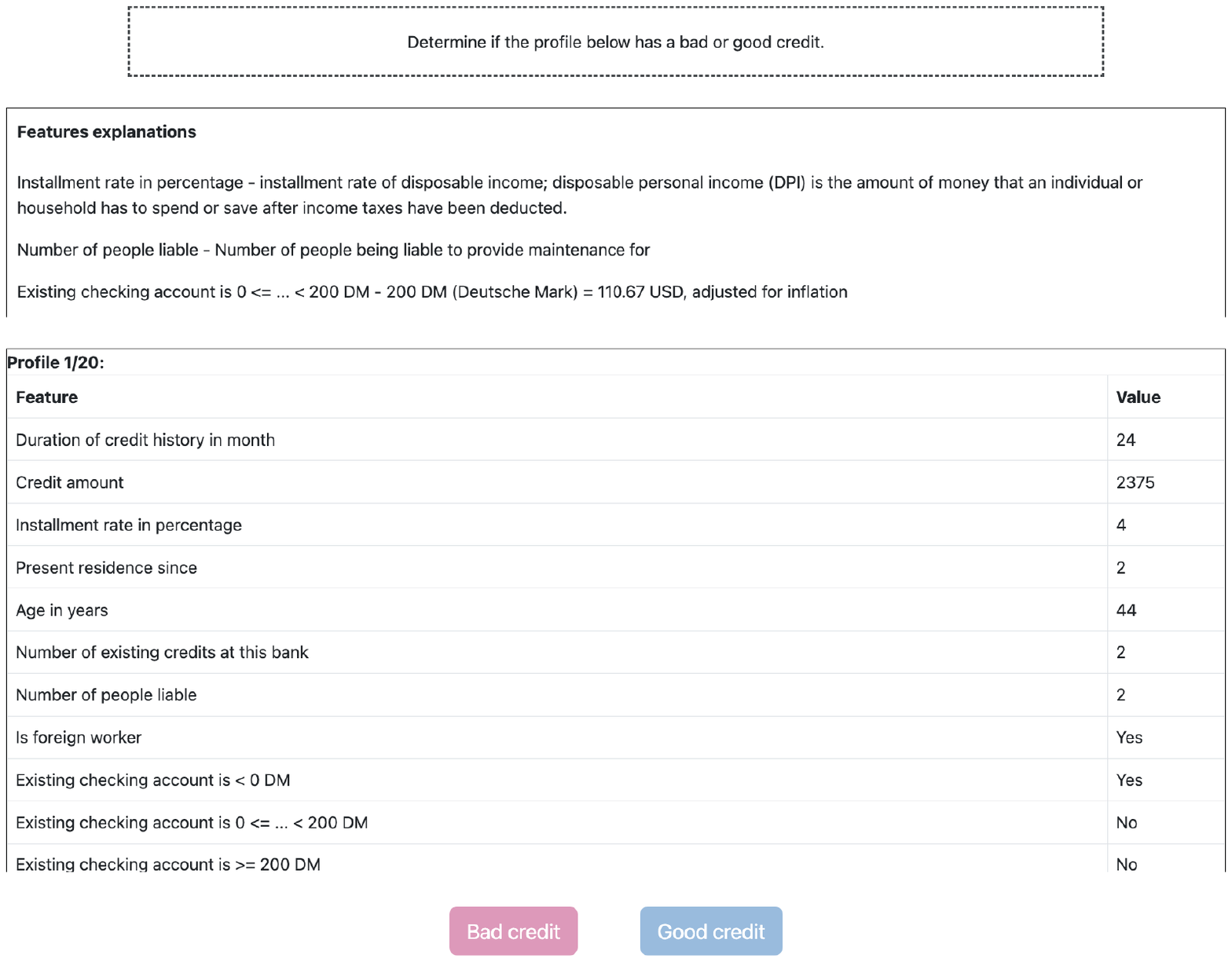}
    \caption{This figure illustrates the task page for the control data feature only condition (\textbf{F}) on the German Credit dataset. There are two main components on this page, the features explanations box and the profile table. The features explanation box has more information on features that might be difficult to understand based on a short description. The profile in the table shows the information on the respective profile the user is required to predict.}
    \label{fig:german_control}
\end{figure*}

\begin{figure*}[t]
    \centering
    \includegraphics[width=\textwidth]{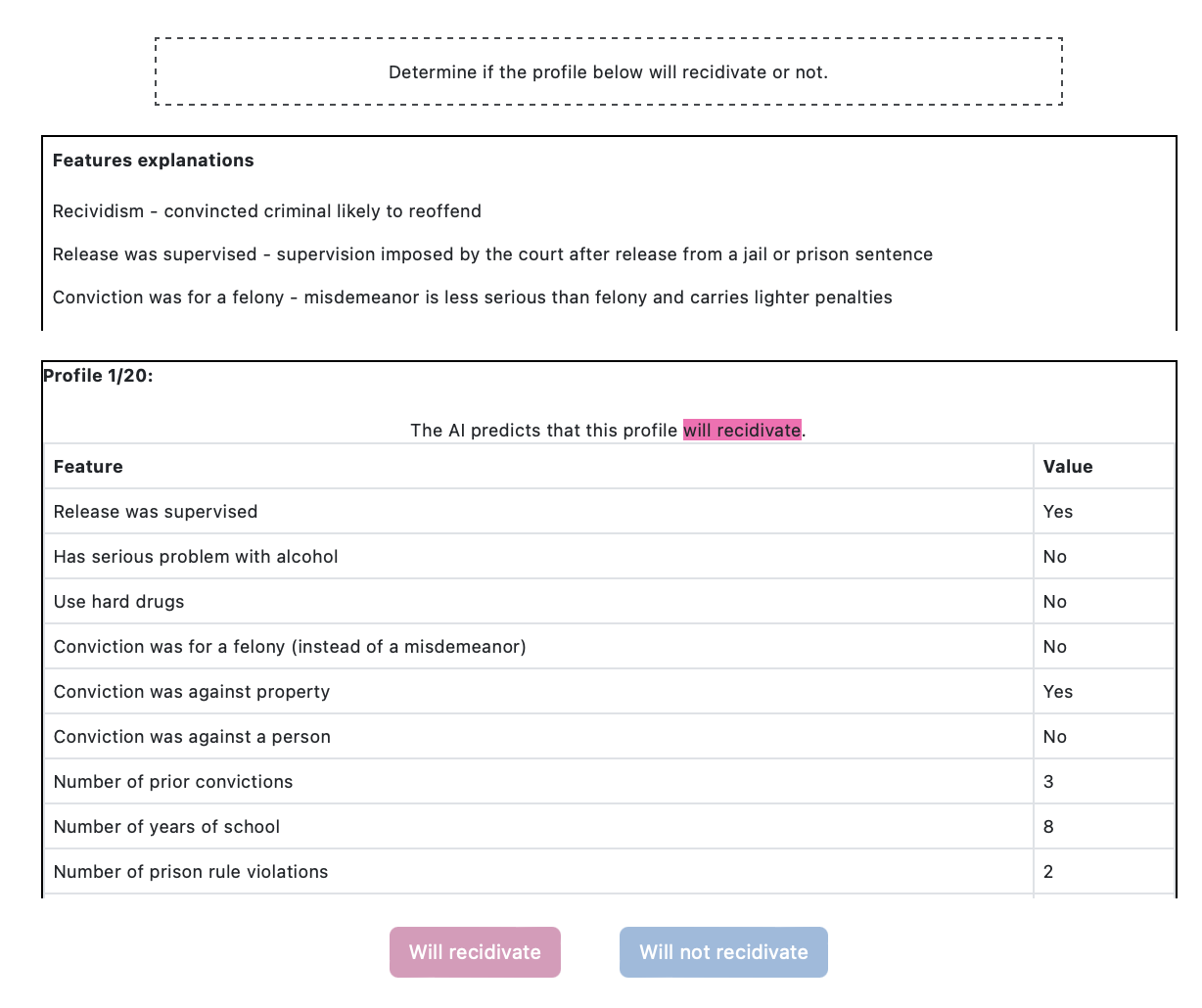}
    \caption{This figure illustrates the task page for the control data feature and model prediction condition (\textbf{FP}) on the RCDV dataset. In addition to the features explanations box and the profile table shown in Figure~\ref{fig:german_control}, there is an additional AI prediction on top of the profile table.}
    \label{fig:rcdv_control}
\end{figure*}

We further show the task page interface designs for the control conditions in Figure~\ref{fig:german_control} and~\ref{fig:rcdv_control}.

\section{Discussion on Measuring Understanding and Trust in XAI}

To gauge people's understanding of XAI methods, the design of survey questions should prioritize qualitative measures that delve into the depths of users' perceptions. Empirical studies include asking participants to directly rate their \textit{understanding} of the AI~\cite{buccinca2020proxy,anik2021data,yang2020visual,lucic2020does,smithrenner2020,binns2018s,wang2021explanations,cheng2019explaining}. However, a mere understanding can be multi-faceted and might manifest as \textit{confidence in understanding}~\cite{kulesza2012tell}, \textit{ease of understanding}~\cite{guo2019visualizing,poursabzi2018manipulating}, or even \textit{confidence in simulation}~\cite{alqaraawi2020evaluating,nguyen2018comparing}. To derive richer insights into users' comprehension of XAI methods, survey designs should also factor in how participants perceive the \textit{intuitiveness}~\cite{szymanski2021visual} and \textit{transparency}~\cite{tsai2021exploring,rader2018explanations} of the AI system. Only through a well-rounded exploration of these qualitative metrics can one truly assess and capture the essence of users' understanding of XAI.

To measure trust in different XAI methods, researchers should use a combination of objective and subjective evaluation metrics. For subjective metrics, most prior empirical studies rely on direct self-reported measures of trust, such as Q2 in Section~\ref{sec:exp} (``I trust the system's predictions.'')~\cite{abdul_cogam_2020,alqaraawi_evaluating_2020,bucinca_proxy_2020,cheng2019explaining,Chromik2021-un,dietvorst_algorithm_2015,slack_assessing_2019,gero_mental_2020,green_principles_2019,kulesza2012tell,poursabzi2018manipulating,ribeiro2018anchors,smithrenner2020,springer_progressive_2018,tsai2021exploring}. Other studies apply some variation of self-reported trust, such as self-reported agreement or reliance~\cite{Chandrasekaran2018-gi}. For objective metrics, our framework supports measures of over-reliance~\cite{bucinca_trust_2021,bussone_role_2015,wang2021explanations,yang2020visual} and under-reliance~\cite{bussone_role_2015,wang2021explanations,yang2020visual} on model predictions. These metrics capture trust by measuring the proportion of times where the user adopts (or does not adopt) the AI prediction when the model is incorrect (or correct). As with measures of understanding, it is important to gauge trust through a combination of objective metrics and self-reported subjective metrics.

\section{IRB Approval}

This study received Institutional Review Board (IRB) approval from the authors' affiliated institutions. Note that researchers who adopt our framework for their own study must receive IRB approval from their own institution(s). The IRB we received for this study does \emph{not} cover any future studies conducted under the OpenHEXAI framework.  

\end{document}